\documentclass[11pt,a4paper]{article}
\usepackage{amsfonts}
\usepackage{epsfig}
\usepackage{graphicx}
\usepackage{amsmath}
\usepackage{amssymb}
\usepackage{color}
\usepackage{cite}

\newcommand \arctanh{\mathop{\rm arctanh}\nolimits}
\newcommand \arccoth{\mathop{\rm arccoth}\nolimits}

\newcommand \sech{\mathop{\rm sech}\nolimits}

\begin{document}

\title{\bf Transference of kinks between $\mathbb{S}^2$ and $\mathbb{R}^2$ Sigma models}

\author{A. Alonso-Izquierdo$^{(a,b)}$, A.J. Balseyro Sebastian$^{(b)}$ and\\  M. A. Gonzalez Leon$^{(a,b)}$
\\ {\normalsize {\it $^{(a)}$ Departamento de Matematica
Aplicada}, {\it Universidad de Salamanca, SPAIN}}\\ {\normalsize {\it $^{(b)}$ IUFFyM}, {\it Universidad de Salamanca, SPAIN}}}

\date{}

\maketitle

\begin{abstract}

In this paper methods for deforming scalar field theories on Euclidean target spaces, in which new field theories are constructed so that solutions are known, are generalized to the context of Sigma models. In particular, deformations between Sigma models on the plane and on the sphere are considered. Three different examples are presented, where the change in the structure of the kink variety and the energy of the deformed kinks during this procedure are studied.   

\noindent

\end{abstract}


\section{Introduction}

Since solitons, finite energy solutions of non-linear differential equations that travel at constant speed while maintaining the same form \cite{Manton2004,Shnir2018}, were identified in the late 19th century, they have become an active field of research. This is not only due to their interesting mathematical properties, but also because of their numerous applications in physics. Indeed, these play an important role in fields such as optical communication \cite{Ablowitz2022, Agrawall1995, Mollenauer2006, Schneider2004} and allow us, for instance, to describe mechanisms for charge and energy transport in molecular systems \cite{Bazeia1999, Cortijo2007, Davydov1985, Guillamat2022, Yakushevich2004}. On the other hand, they are also present in cosmology and high-energy physics \cite{Kibble1976, Kolb1990, Vachaspati2006, Vilenkin1985, Vilenkin2000, Weinberg2012} or condensed matter physics \cite{Bishop1980, Cheng2021, Eschenfelder1981, Jona1993, Machon2016, Pollard2020, Strukov1998, Vanhaverbeke2008} among others.

In particular, in this work we shall focus on solitons in $1+1-$dimensional scalar field theories, where they are referred to as kinks. While the analytical identification of kinks is generally a difficult task, several methods have been developed to address this challenge. One successful method for finding kinks is the deformation method developed by Bazeia et al \cite{Afonso2007, Almeida2004, AlonsoIzquierdo2013, Bazeia2002, Bazeia2006a, Bazeia2006b, Bazeia2011, Bazeia2013, Bazeia2017, Chumbes2010, Cruz2009}. This method employs a field theory as a seed to construct new field theories for which kinks can be analytically identified. The main idea behind this method is to exploit the transformations that relate different Bogomol'nyi equations in order to connect solutions across various field theories. However, these techniques have thus far only been applied to obtain kinks in field theories with Euclidean target spaces.

In this paper, these techniques are generalized so that deformations of Sigma models are considered. In particular, deformations between Sigma models on the plane and on the sphere $\mathbb{S}^2$ shall be examined. There are two reasons for choosing the sphere as the non-Euclidean target manifold of the deformed Sigma models. First, unlike the plane, the sphere is a compact manifold. This opens up the possibility of significant changes in the topological sectors of deformed kinks.  The second reason is that constructing new Sigma models on the sphere, see for example \cite{Alonso2010,Alonso2018}, has potential applications in fields such as spintronics \cite{Chumak2015, Hirohata2014, Hirohata2020, Kimura2012, Lesne2016, Serga2010}. Indeed, Sigma models on the sphere can be employed to describe spin, which in this field is used to process and store information.

The outline of the paper is as follows.  In Section $2$ we provide a brief summary of Sigma models and how Bogomol'nyi equations arise in this context. Section $3$ introduces the procedure for transferring kinks between models on the plane and the sphere. In Section $4$, two different coordinates on the sphere are employed to construct three representative examples of deformations. Lastly, conclusions are drawn in the last section.

\section{Kinks in non-linear Sigma models}

Let us consider a field theory on the $(1+1)-$dimensional Minkowski space with an $n-$dimensional Riemannian manifold $M$ as target space. Fixing a chart on $M$ defines real scalar fields $\phi^i:\mathbb{R}^{1,1}\rightarrow \mathbb{R}$, which will be collectively denoted as $\phi=(\phi^1,\dots,\phi^n)$. Let us consider the Sigma model on $M$ defined by the action functional:
\begin{equation}
S[\phi]= \int_{\mathbb{R}^{1,1}}  \, \left[ \frac{1}{2} \eta^{\mu \nu} g_{ij}\frac{\partial\phi^i}{\partial x^\nu}\frac{\partial\phi^j}{\partial x^\mu} -V(\phi)\right] \, dx ~ dt \, , \label{action2}
\end{equation}
where $x^{\mu}$ denote the coordinates in $\mathbb{R}^{1,1}$ with $(x^0,x^1)=(t,x)$, the Minkowski metric has been chosen as $\eta_{\mu\nu}={\rm diag} \, (1,-1)$, we denote as $g_{ij}(\phi)$ the components of the metric tensor in the chart on $M$, the non-negative function $V:M\rightarrow \mathbb{R}$ is the potential and where Einstein's summation convention of indices is employed. The Euler-Lagrange equations derived from (\ref{action2}) are the non-linear PDEs
\begin{equation}
\eta^{\mu\nu} \frac{\partial^2 \phi^i}{\partial x^\mu  \partial x^\nu} + \eta^{\mu\nu}\Gamma_{jk}^i \frac{\partial \phi^j}{\partial x^\mu} \frac{\partial \phi^k}{\partial x^\nu}+ g^{ij} \frac{\partial V}{\partial \phi^j}=0  \, \quad i,j=1,\dots, n \, , \label{fenlv2}
\end{equation}
where $\Gamma_{jk}^i$ are the Christoffel symbols corresponding to $g$. Invariance under time translations in the action functional (\ref{action2}) implies the conservation of the total energy
\begin{equation*}
E[\phi]=\int^{\infty}_{-\infty}  \, \varepsilon(x,t) \, dx= \int^{\infty}_{-\infty}  \, \left[ \frac{1}{2} g_{ij}\left(\frac{\partial\phi^i}{\partial t}\frac{\partial\phi^j}{\partial t}+\frac{\partial\phi^i}{\partial x}\frac{\partial\phi^j}{\partial x}\right) + V(\phi)\right] \, dx
\end{equation*}
for any solution of equations (\ref{fenlv2}), where the energy density has been denoted as $\varepsilon(x,t)$. The configuration space ${\cal C}$ is comprised of finite energy configurations
\[
{\cal C} =\{ \phi(t,x) \in M \, |\, \, E[\phi(t,x)] < + \infty \} \, .
\]
Consequently, the elements of ${\cal C}$ must comply with the asymptotic conditions
\[
\lim_{x\rightarrow \pm \infty} \frac{\partial \phi^i(t,x)}{\partial t} = \lim_{x\rightarrow \pm \infty} \frac{\partial \phi^i(t,x)}{\partial x} =0  \, \, \forall i \, ,\hspace{1cm}  \lim_{x\rightarrow \pm \infty}  \phi(t,x) \in {\cal M} \, ,
\]
where ${\cal M}$ is the set of zeroes of the potential function $V(\phi)$, also referred to as vacua of the model:
\[
{\cal M} = \{v_j \in M \, |\, \,  V(v_j) =0, \,\, j=1,2,\dots \}\, ,
\]
which in principle is assumed to be discrete. In order to identify kinks, static solutions of (\ref{fenlv2}) will be sought, allowing the retrieval of traveling solutions by applying a boost. This simplifies the search for kinks, which now does not involve solving equations in partial derivatives as only ordinary derivatives of fields respect to $x$ remain. On the other hand, if the potential $V$ can be written in terms of a superpotential $W:M\rightarrow\mathbb{R}$ which is asssumed to be differentiable 
\begin{equation}
V(\phi)= \frac{1}{2} g^{ij}\, \frac{\partial W}{\partial \phi^i} \frac{\partial W}{\partial \phi^j}  \, \, \, , \, \, \, i,j=1,\dots, n \, ,  \label{ps}
\end{equation}
then the Bogomol'nyi arrangement \cite{Bogomolny1976} allows us to write the energy functional as follows
\[
E[\phi]=\frac{1}{2}\int dx \,\left(g_{ij} \left[\frac{d\phi^i}{dx}+ (-1)^{\epsilon} \, g^{mi} \frac{\partial W}{\partial \phi^m}\right] \left[\frac{d\phi^j}{dx}+(-1)^{\epsilon} \, g^{nj} \frac{\partial W}{\partial \phi^n}\right]\right) + T \, ,
\]
where $\epsilon=0, 1$ and $T$ is defined as
\begin{equation}
T=\left\vert \int dx \,   \frac{d\phi^i}{dx}\frac{\partial W}{\partial \phi^i} \, \,\right\vert \, . \label{charge}
\end{equation}
This arrangement in the static energy $E[\phi]$ allows us to easily identify first order differential equations for solutions that minimize this functional in the configuration space ${\cal C}$ within a topological sector. These solutions, referred to as BPS kinks in the literature, must satisfy then the system of first order differential equations
\begin{equation}
\frac{d\phi^i}{dx}=(-1)^{\epsilon} \, g^{ij} \, \frac{\partial W}{\partial \phi^j} \, ,\, \, \, \, i,j=1,\dots, n. \label{bpsb}
\end{equation}
The role of the parameter $\epsilon$ is to distinguish between kinks and antinkinks, as it can be trivially checked by making the spatial variable $x$ absorb the global factor $(-1)^{\epsilon}$. Moreover, the magnitude $T$ defined in (\ref{charge}) becomes a topological charge and coincides with the energy of the static kink
\begin{equation}\label{eq:EnergyBPS}
   T=\Big\vert \lim_{x\rightarrow \infty} W[\phi(x)]- \lim_{x\rightarrow -\infty}W[\phi(x)]\Big\vert \, ,
\end{equation}
Notice that this quantity only depends on the evaluation of the superpotential at the initial and final vacua, which are asymptotically connected by BPS kinks. In summary, for potentials of the form \eqref{ps} Bogomol'nyi equations \eqref{bpsb} allow us to find BPS kinks, which are also static solutions of \eqref{fenlv2} that minimize the static energy of the field theory.

\section{Tranference between Sigma models on the plane and the sphere}

Let us consider first the Euclidean plane $(\mathbb{R}^2,\delta_{ij})$ as the target manifold of a model that admits a superpotential. When Cartesian coordinates $\{\phi^1,\phi^2\}$ are chosen in the plane, first-order differential equations (\ref{bpsb}) can be written in terms of the superpotential $W(\phi^1,\phi^2)$ as follows

\begin{equation}
\frac{d\phi^i}{dx}= (-1)^{\epsilon} \frac{\partial W}{\partial \phi^i}\  ,\quad i=1,2.\label{foeq1}
\end{equation}
Solutions of these equations correspond to kinks for the model with potential function \eqref{ps} on the plane, which now reads
\[
V(\phi^1,\phi^2) = \frac{1}{2} \left[  \left( \frac{\partial W}{\partial \phi^1}\right)^2+ \left( \frac{\partial W}{\partial \phi^2}\right)^2\right]\,.
\]

Let $\Sigma(t,x) \equiv (\phi^1(t,x), \phi^2(t,x))$ be a kink-type solution of equations (\ref{foeq1}) in the plane. On the other hand, let us consider a model on the sphere $(\mathbb{S}^2,g)$ with the metric tensor $g$ inherited by embedding it in $\mathbb{R}^3$. In order to construct this model on the sphere with potential $\widetilde{V}$, let us denote as $\widetilde{W}$ the superpotential on the sphere and the coordinates on the sphere as $\left\{\psi^1,\psi^2\right\}$. The Bogomol'nyi arrangement allows us to derive the Bogomol'nyi equations for the model on the sphere
\begin{equation}
\frac{d\psi^i}{d\tilde{x}}=(-1)^{\epsilon}\, g^{ij} \frac{\partial \widetilde{W}}{\partial \psi^j}\,,\label{foeq22}
\end{equation}
where $\tilde{x}$ denotes the spatial coordinate in this model. Deformation methods are a collection of techniques in the literature that can be employed to construct new field theories for which solutions can be identified from the previous ones \cite{Afonso2007, Almeida2004, AlonsoIzquierdo2013, Bazeia2002, Bazeia2006a, Bazeia2006b, Bazeia2011, Bazeia2013, Bazeia2017, Chumbes2010, Cruz2009}. In particular, this procedure  takes as its starting point a model with known solutions. Then, new models are constructed mapping the original solutions to the new target space. If one attempts to deform a model in the plane sending solutions to a model in the sphere, one must specify how curves are transferred by defining a relation between coordinates on both charts $\psi^i(\phi^1,\phi^2)$ for $i=1,2$ and how both parametrizations of curves are related $x \rightarrow \tilde{x}$. More precisely, in this work we shall be focusing on symmetrical dilations $\psi^i=\mu \, \phi^i$ with $\mu>0$. Notice that both the coordinates on the sphere and the dilation must be chosen so that the solutions in the plane that are transferred via this identification of coordinates are well-defined in the chart on the sphere. Additionally, when the reparametrization $\tilde{x}(x)$ is invertible, curves on the sphere can be explicitly written for each coordinate $i=1,2$ as
$$\psi^i(\tilde{x})=\mu \, \phi^i(x(\tilde{x})) \,.$$ 
 Moreover, let us consider coordinates for which the sphere is locally conformally flat, i.e. $g_{ij}=G(\psi)\delta_{ij}$ with a positive definite function $G(\psi)$. In this scenario, if the reparametrization of curves is defined as follows 
\begin{equation}\label{eq:GeneralReparemetrisation}
    \frac{d\tilde{x}}{dx}=\frac{1}{G(\mu \, \phi(x))}\,,
\end{equation}
then a superpotential on the sphere for which $\psi(\tilde{x})$ is a solution of the new Bogomol'nyi equations \eqref{foeq22} can be straightforwardly found  
\begin{equation}\label{eq:RelationSuperpotentials}
    \widetilde{W}(\psi^1,\psi^2)=\mu^2 W\left(\frac{\psi^1}{\mu},\frac{\psi^2}{\mu}\right)\,.
\end{equation}
 This leads to a potential function of a model in the sphere for which these deformed curves $\psi(\tilde{x})$ are static solutions
\begin{equation}\label{eq:PotentialRelation} 
    \widetilde{V}(\psi)=\frac{\mu^2}{G(\psi)}V\left(\frac{\psi}{\mu}\right) \,.
\end{equation}
It is worth noticing that conformal transformations ensure that the new orbit flow equation on the sphere remains identical in form to the original in the plane. In consequence, orbits in both charts, on the plane and on the sphere, will be formally identical up to a dilation. In addition to this, the fact that the transference is given by a symmetric dilation allows us to identify a relation between the energy of the original and the transferred kinks. It is straightforward to check that the energy of the transferred kinks is related to those in the original model by a factor that depends on the dilation parameter
\begin{equation}\label{eq:DilatacionEnergia}
     \widetilde{E}[\psi]=\mu^2 \, E[\phi] \,.
\end{equation}
This is, the dilation allows us to modulate the energy of the deformed kinks. Alternatively, this result can be obtained by noticing that the factor $\mu^2$ appears in the potential function $\widetilde{V}$, modulating its peaks and therefore the energy of the corresponding kink. In summary, the knowledge of a kink solution $\phi_K(x)$ in a  Sigma model with the Euclidean plane as target space implies the knowledge of kink solutions $\psi_K(\tilde{x})$ for Sigma models with the sphere as target manifold. 

Lastly, without the risk of confusing notation, all indices of coordinates will henceforth be written as subscripts in order to alleviate notation, without any implication in the change of contravariance or covariance.

\section{Transference by stereographic and Mercator projections}

In order to illustrate the transference of solutions between models in the plane and the sphere, let us consider two coordinate systems on the sphere for which the sphere is locally conformally flat. These shall be the stereographic and the Mercator projections, briefly summarized in Appendix \ref{sec:appendixA}. This procedure allows us to transfer, employing any of these conformal coordinates and a properly defined reparametrization of curves, static solutions of a model in the plane to static solutions of other Sigma models on the sphere. Indeed, the superpotential \eqref{eq:RelationSuperpotentials} for the new field theory can always be found. 

By construction, in the stereographic projection the north pole $N$ in the sphere is excluded.  This implies that this point is not accessible by the transference of solutions from the plane to the sphere. On the other hand, in the Mercator projection two points $\left\{N,S\right\}$ in the sphere are excluded, the north and the south poles. In this case two points are not accessible. Let us denote these ``singular projection points'' as $s_i$. Even if these points are not accessible by the antiprojections, these may be contained in deformed solutions when the original solutions in the plane tend to the infinities that are being mapped to precisely these points. For global factors $G(\psi)$ of the metric for which the limits at singular projection points
\begin{equation}
    \widetilde{V}(s_i)= \displaystyle\lim_{\psi\rightarrow s_i}\frac{\mu^2}{G(\psi)}V\left(\frac{\psi}{\mu}\right) \,
\end{equation}
exist, different interesting scenarios will be studied in the examples presented in this section. Conversely, the process of transference of solutions from the sphere to the plane could also be considered, where by construction the singular projection points are not sent to the plane. This would also imply the possibility of a change in the topological sectors of the resulting kink variety in the plane.

\subsection{Example A}
\label{sec:ExampleA}
\begin{figure}[ht]
\centerline{\includegraphics[height=3cm]{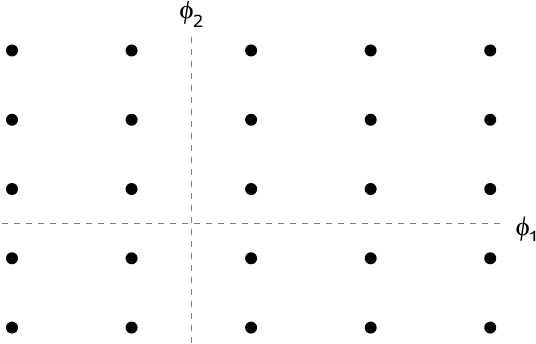} \hspace{0.3cm}
\includegraphics[height=3cm]{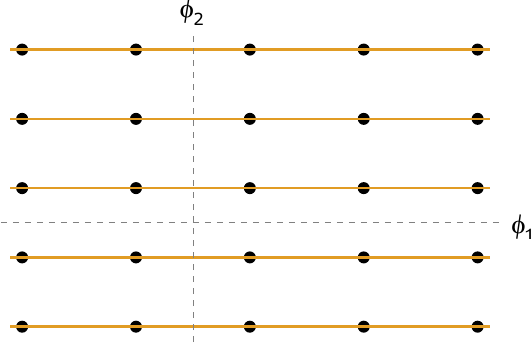}  \hspace{0.3cm} \includegraphics[height=3cm]{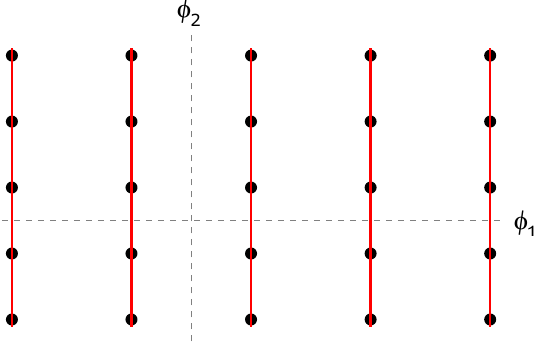}}
\caption{\small Vacua of the sine-Gordon model in the plane are shown on the left and singular $\Phi_1$ and $\Phi_2-$kinks are represented in the center and on the right respectively. These singular kinks correspond to vertical and horizontal segments replicated in the plane.}\label{figure:SGR2sing}
\end{figure}

\begin{figure}[ht]
\centerline{\includegraphics[height=4.5cm]{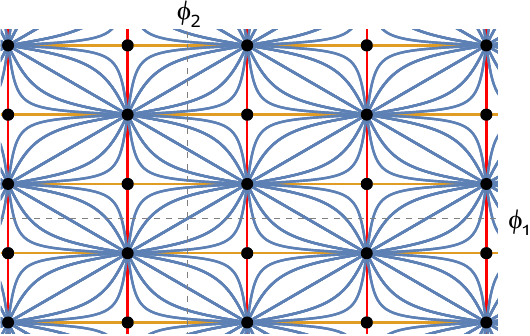} \hspace{0.5cm}
\includegraphics[height=4.5cm]{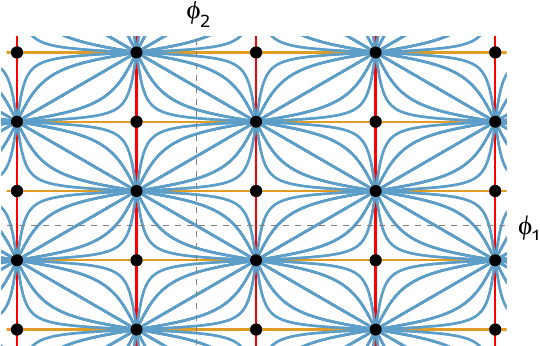}}
\caption{\small Two families of kinks appear in each cell delimited by singular kinks in the plane, that with $\epsilon_1+\epsilon_2=0$ mod $2$ (left) and that with $\epsilon_1+\epsilon_2=1$ mod $2$ (right). These members $\Sigma(x)$ correspond in limits of constants $x_{0,1}$ and $x_{0,1}$ at infinities to both types of singular kinks.}\label{figure:SGR2fam}
\end{figure}

\begin{figure}[ht]
\centerline{\includegraphics[height=4cm]{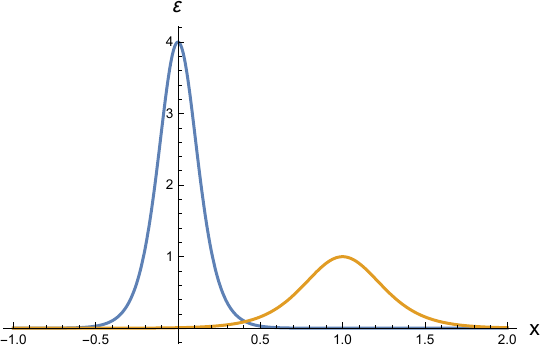}  \hspace{0.3cm} \includegraphics[height=4cm]{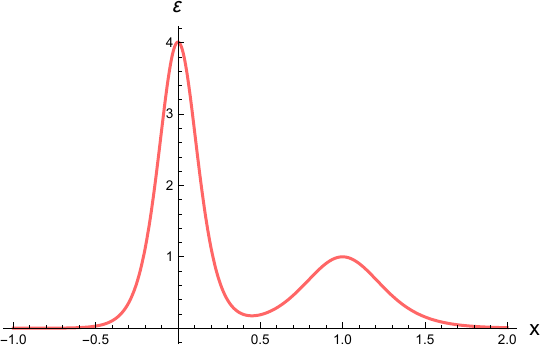}}
\caption{\small Energy densities for singular kinks are represented on the left and for members of the family of solutions on the right. Values of parameters $\alpha_1=2$, $\alpha_2=1$, $x_{0,1}=0$ and $x_{0,2}=1$ have been chosen. While singular kinks exhibit one peak, two appear for members of the family of kinks.}\label{figure:SGR2Densidades}
\end{figure}

 As first example let us consider the sine-Gordon model in the plane. In particular, let us consider the model in the plane with action
\begin{eqnarray*}
S[\phi]&=& \int_{\mathbb{R}^{1,1}}  \, \left[ \frac{1}{2}\left(\left(\frac{\partial\phi_1}{\partial t}\right)^2+ \left(\frac{\partial\phi_2}{\partial t}\right)^2-\left(\frac{\partial\phi_1}{\partial x}\right)^2 -\left(\frac{\partial\phi_2}{\partial x}\right)^2\right) - V(\phi)\right] \, dx \, dt
\end{eqnarray*}
 where the potential function is chosen to depend on two parameters $\alpha_1,\alpha_2\in\mathbb{R}$ and is of the form
 \begin{equation}\label{eq:PotentialPlaneGordon}
V(\phi_1,\phi_2)= \frac{\alpha_1^2}{2}  \cos^2 \left(\pi\phi_1\right)+\frac{\alpha_2^2}{2}  \cos^2 \left(\pi \phi_2\right).
\end{equation}
 This model admits two non-equivalent superpotentials ditinguished by a relative minus sign
\[
W^{(\pm)}(\phi_1,\phi_2) = \frac{\alpha_1}{\pi} \sin \left(\pi \phi_1\right) \pm \frac{\alpha_2}{\pi} \sin \left( \pi \phi_2\right),
\]
which allows the Bogomol'nyi arrangement. Now, the set of vacua ${\cal M}$ of this model consists of an infinite bidimensional lattice of points
\begin{eqnarray*}
{\cal M}& = & \textstyle \left\{ v^{n_1 n_2}=\left(\frac{2n_1+1}{2}, \frac{2n_2+1}{2}\right)/ n_1,n_2\in \mathbb{Z}\right\} \, ,
\end{eqnarray*}
 see Figure \ref{figure:SGR2sing} (left). Solutions of Bogomol'nyi equations will be classified into singular kinks, where one of the coordinates remains constant along the orbit, and into families of kinks. In order to alleviate notation, let us introduce an auxiliary function written in terms of the Gudermannian function
 \begin{equation*}
     f_i(x)=(n_i+1) + \frac{1}{\pi} {\rm Gd} \left[ (-1)^{\epsilon_i} \pi \alpha_i (x-x_{0,i})\right]\,, \quad \text{where} \quad {\rm Gd}\,[y]=-\frac{\pi}{2}+2\arctan \text{e}^{y}\,,
 \end{equation*}
for $i=0,1$, $\epsilon_i=0,1$ and $x_{0,i}\in\mathbb{R}$. A classification of solutions is now presented:

\begin{itemize}
    \item \textbf{Singular $\Phi_1-$kinks}: When the trial orbit $\phi_2=\frac{2n_2+1}{2}$ with $n_2\in\mathbb{Z}$ is imposed in Bogomol'nyi equations one obtains a solution, an energy density and an energy for all integers $n_1,n_2$ 
    \begin{equation*}
        \Phi_1(x)=\left(f_1(x),\frac{2n_2+1}{2} \right) , \qquad \varepsilon(x)=\alpha_1^2 \sech^2{\left[\alpha_1 \pi (-1)^{\epsilon_1}(x-x_{0,1})\right]} \,,\qquad E[\Phi_1]=\frac{2|\alpha_1|}{\pi}\,.
    \end{equation*}
    These singular kinks describe horizontal segments replicated in all the plane, see Figure \ref{figure:SGR2sing} (center).
    
    \item \textbf{Singular $\Phi_2-$kinks}: When a trial orbit where now the coordinate $\phi_1=\frac{2n_1+1}{2}$ with $n_2\in\mathbb{Z}$ remains constant is imposed, the following solution is obtained once again $\forall n_1,n_2\in \mathbb{Z}$:
    \begin{equation*}
        \Phi_2(x)=\left(\frac{2n_1+1}{2} , f_2(x)\right) ,\qquad \varepsilon(x)=\alpha_2^2 \sech^2{\left[\alpha_2 \pi (-1)^{\epsilon_2}(x-x_{0,2})\right]} \,, \qquad E[\Phi_2]=\frac{2|\alpha_2|}{\pi}\,.
    \end{equation*}
    These singular kinks describe now vertical segments, see Figure \ref{figure:SGR2sing} (right).
    
    \item \textbf{Families of kinks $\Sigma$}: Bogomol'nyi equations can be solved in this case in a general form, leading to two families of kinks $\forall n_1,n_2\in \mathbb{Z}$
    \begin{equation*}
        \Sigma(x)=\left(f_1(x), f_2(x)\right) ,  \quad \varepsilon(x)=\sum_{i=1}^2\alpha_i^2 \sech^2{\left[\alpha_i \pi (-1)^{\epsilon_i}(x-x_{0,i})\right]} \,,\quad  E[\Sigma]=\frac{2\left(|\alpha_1|+|\alpha_2|\right)}{\pi}\,,
    \end{equation*}
    depending on both $\epsilon_1$ and $\epsilon_2$. Both families of kinks densely fill each cell in the plane, see Figure \ref{figure:SGR2fam}.
\end{itemize}
Notice that all possible integers $n_1$ and $n_2$ generate all possible solutions between adjacent vacua. Moreover, while the energy density profile of singular kinks is comprised by one peak, two appear for the family of kinks, see Figure \ref{figure:SGR2Densidades}.


 It is worth noting that parameters $\alpha_1$ and $\alpha_2$ control the energy of the kinks in the plane. Moreover, regardless of the signs of $\alpha_1$ and $\alpha_2$ the energy sum rule $E[\Sigma]=E[\Phi_1]+E[\Phi_2]$ arises. Of course, this is consistent with limits of the families of kinks (solution $\Sigma$). Indeed, one of the components of $\Sigma(x)$ tends in the corresponding limit $x_{0,i}\rightarrow \pm\infty$ to the value of the component of a vacuum point it is joining. 

\begin{figure}[ht]
\centerline{\includegraphics[height=5cm]{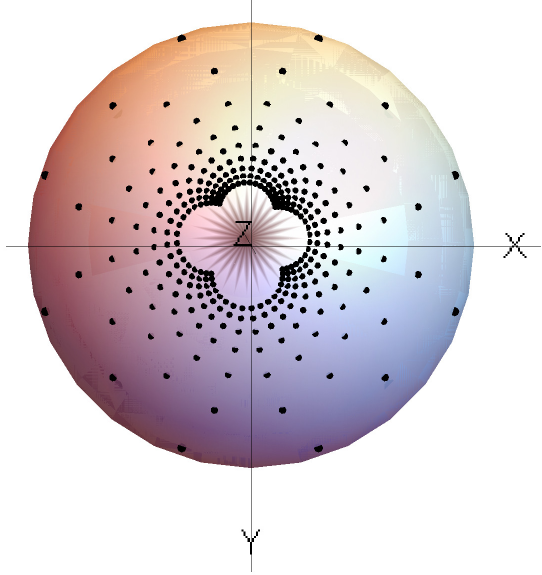} \hspace{0.3cm}
\includegraphics[height=5cm]{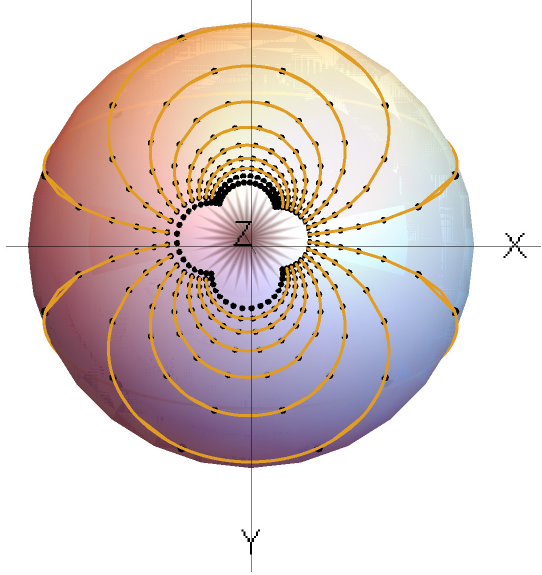}  \hspace{0.3cm} \includegraphics[height=5cm]{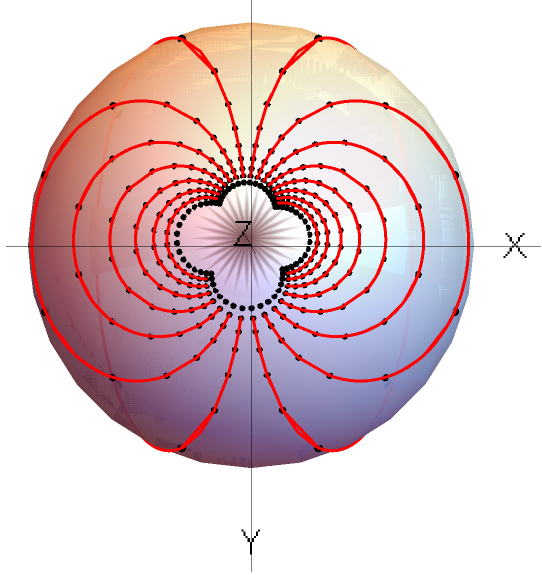}}
\caption{\small Transferred vacua (left), transferred singular $\Phi_1-$kinks (center) and transferred singular $\Phi_2-$kinks (right) from the sine-Gordon model \eqref{eq:PotentialPlaneGordon} in the plane to the Sigma model \eqref{eq:Mod1PotEsf} on the sphere. An infinite number of vacua concentrate around the north pole, without being able to reach it because the potential is infinite at this point.}\label{figure:SGS2sing}
\end{figure}

\begin{figure}[ht]
\centerline{\includegraphics[height=5cm]{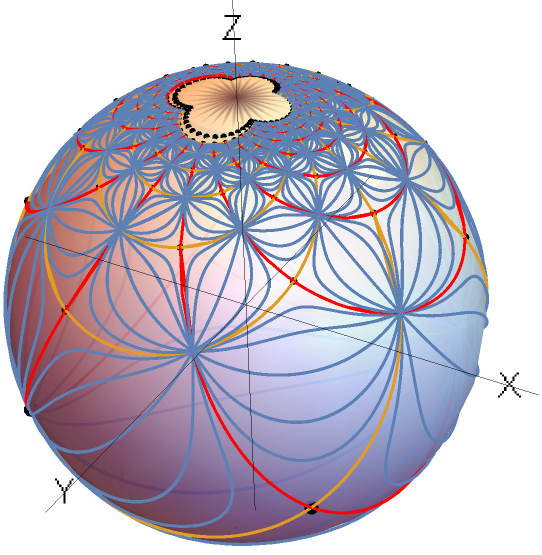} \hspace{0.5cm}
\includegraphics[height=5cm]{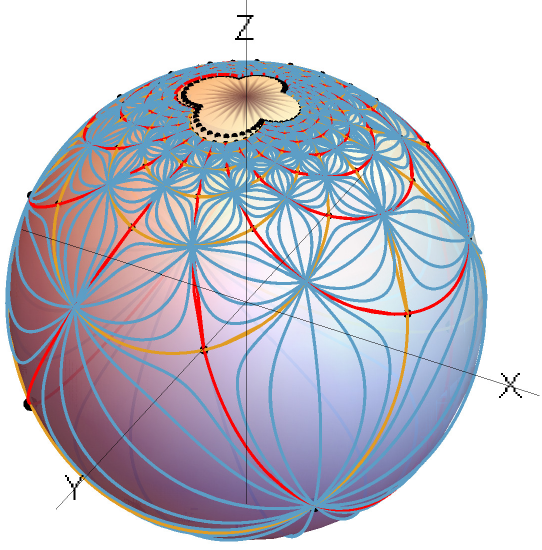}}
\caption{\small The two families of solutions transferred from model \eqref{eq:PotentialPlaneGordon} in the plane to the one on the sphere \eqref{eq:Mod1PotEsf} are depicted. These two families of kinks appear joining different vacua in each region delimited by the transferred singular kinks. Notice that while the distance between adjacent cells for the original solutions in the plane is fixed, on the sphere this distance tends to zero as solutions approach the north pole, which can never be reached.}\label{figure:SGS2fam}
\end{figure}

On the other hand, let us consider a model in the sphere written in stereographic coordinates $\left\{\psi_1,\psi_2\right\}$ with action
\begin{equation}\label{eq:actionMod1Esfera}
\widetilde{S}[\psi]= \int_{\mathbb{R}^{1,1}}  \, \left[ \frac{2}{\left(1+\psi_1^2+\psi_2^2\right)^2}\left(\left(\frac{\partial\psi_1}{\partial \tilde{t}}\right)^2+ \left(\frac{\partial\psi_2}{\partial \tilde{t}}\right)^2-\left(\frac{\partial\psi_1}{\partial \tilde{x}}\right)^2 -\left(\frac{\partial\psi_2}{\partial \tilde{x}}\right)^2\right) - \widetilde{V}(\psi)\right] \, d\tilde{x} \, d\tilde{t} \,.
\end{equation}
The transference of solutions of the Sigma model in the plane to the sphere can be performed by the previously described procedure via the stereographic antiprojection. This leads to a Sigma model on the sphere with a superpotential of the form \eqref{eq:RelationSuperpotentials}, now in terms of stereographic coordinates on the sphere
\[
\widetilde{W}(\psi_1,\psi_2) =  \frac{ \mu^2 \alpha_1}{\pi} \sin \left( \frac{\pi \psi_1}{\mu}\right) +\frac{\mu^2 \alpha_2}{\pi} \sin \left(\frac{\pi \psi_2}{\mu}\right)\,.
\]
The deformed potential on the sphere $\widetilde{V}(\psi)$ can be constructed from this superpotential, which presents a global factor modulated by both a metric factor and the parameter of the dilation 
\begin{equation}\label{eq:Mod1PotEsf}
    \widetilde{V}(\psi_1,\psi_2)= \mu^2 \, \frac{\left(1+\psi^2_1+\psi^2_2 \right)^{2}}{8} \left(\alpha_1^2  \cos^2 \left( \frac{\pi\psi_1}{\mu}\right)+\alpha_2^2  \cos^2 \left( \frac{\pi \psi_2}{\mu}\right)\right)\,.
\end{equation}
This potential also presents an infinite number of vacua. Indeed, coordinates of vacua in the stereographic plane are a dilation of the coordinates of the vacua in the Euclidean plane. However, since infinities in the stereographic plane correspond to the north pole in the sphere, an infinite number of vacua concentrate around this point, see Figure \ref{figure:SGS2sing} (left). Since in these coordinates the sphere is locally conformally flat, kink orbits in the stereographic plane will be analytically identical to those in the plane. The reparametrization of each transferred solution, on the other hand, will depend on the explicit form of the solution in the plane $\phi(x)=\left(\phi_1(x),\phi_2(x)\right)$ which is being transferred. Indeed, by construction the reparametrization \eqref{eq:GeneralReparemetrisation}, which in this case reads 
\begin{equation}
    \tilde{x}=\tilde{x}_0+\displaystyle\int\frac{4}{\left(1+\mu^2 \phi_1^2(x)+\mu^2 \phi_2^2(x)\right)^2} \, dx\,,
\end{equation}
with $\tilde{x}_0\in\mathbb{R}$, allows us to identify solutions $\psi^i(\tilde{x})=\mu \, \phi^i(x(\tilde{x}))$ of the new Bogomol'nyi equations, describing now trajectories in the sphere, see Figures \ref{figure:SGS2sing} and \ref{figure:SGS2fam}. On one hand, orbits of the transferred singular kinks, which are segments in the stereographic plane, become curve on the sphere. Similarly, members of the family of kinks in each of these cells are curved accordingly to asymptotically connect the transferred vacua. It is worth highlighting that the same energy sum rules as before hold since all deformed kinks are rescaled by the same factor \eqref{eq:DilatacionEnergia}. 

 Even though the potential in the plane \eqref{eq:PotentialPlaneGordon} is bounded, this new potential on the sphere diverges at the north pole. Consequently, the singularity of the potential excludes this projection singular point from the set of accessible points of the target manifold.

\subsection{Example B}

 Let us consider as second example a model in the plane constructed from the following rational superpotential written in Cartesian coordinates
\begin{equation}\label{eq:SuperpotentialModel2}
	W(\phi_1,\phi_2)=\frac{a_1 \phi_1 + a_2 \phi_2}{b^2 +\phi^2_1+\phi_2^2} \, ,
\end{equation}
with arbitrary parameters  $b,a_1,a_2\in\mathbb{R}$. This superpotential engenders a potential function on the plane of the form
\begin{equation}\label{eq:PotentialModel2PlaneGeneral}
	V(\phi_1,\phi_2)=\frac{1}{2}\frac{\left[a_2 \left(b^2+\phi^2_1-\phi_2^2\right)-2a_1 \phi_1 \phi_2\right]^2+ \left[a_1 \left(b^2-\phi^2_1+\phi_2^2\right)-2a_2 \phi_1 \phi_2\right]^2}{\left(b^2+\phi^2_1+\phi_2^2\right)^4}\,.
\end{equation}
 It is worth noting that unlike the model in the plane of the previous section, this potential is well-defined in the limit $\phi_1^2+\phi_2^2\rightarrow\infty$. In particular, this potential presents, on one hand two vacua whose position is modulated by parameters $a_1$ and $a_2$, and on the other tends to zero at all points at infinities, which shall be denoted as $v_{\infty}$: 
\begin{equation}
	\mathcal{M}=\left\{\left(\frac{a_1 b}{\sqrt{a_1^2+a_2^2}},\frac{a_2 b}{\sqrt{a_1^2+a_2^2}}\right),\left(\frac{-a_1 b}{\sqrt{a_1^2+a_2^2}},\frac{-a_2 b}{\sqrt{a_1^2+a_2^2}}\right),v_{\infty}\right\} \,.
\end{equation}
This scenario, involving this particular configuration of vacua at the infinities, is interesting as all these points $v_{\infty}$ will be mapped to the same point on the sphere, specifically the north pole. Moreover, making use of polar coordinates in the parameter plane $a_1=\lambda \cos{\nu}$, $a_2=\lambda \sin{\nu}$, the set of vacua can be written as follows
\begin{equation}
	\mathcal{M}=\left\{\left( b \cos\nu, b \sin{\nu}\right),\left( b \cos(\nu+\pi), b \sin{(\nu+\pi)}\right),v_{\infty}\right\} \,.
\end{equation}
  This indicates that each angle value $\nu\in[0,\pi)$ defines the position of the two isolated vacua, which are antipodal points in a circumference of radius $b$. On the other hand, Bogomol'nyi equations for this superpotential are given by
\begin{align}
	& \frac{d \phi_1}{dx}=(-1)^{\epsilon}  ~ \frac{a_1 \left(b^2-\phi^2_1+\phi_2^2\right)-2 a_2 \phi_1 \phi_2 }{\left(b^2+\phi^2_1+\phi_2^2\right)^2} \, , \label{eq:FirstOrderEqsModelo2a}\\
	& \frac{d \phi_2}{dx}=(-1)^{\epsilon}  ~ \frac{a_2 \left(b^2+\phi^2_1-\phi_2^2\right)-2 a_1 \phi_1 \phi_2}{\left(b^2+\phi^2_1+\phi_2^2\right)^2} \,, \label{eq:FirstOrderEqsModelo2b}
\end{align}
where once more $\epsilon=0,1$. The orbit flow equation, which is given by
\begin{equation}\label{eq:OrbitEquationModel2}
	\frac{d\phi_2}{d\phi_1}=\frac{a_2 (b^2+\phi^2_1-\phi_2^2)-2 a_1 \phi_1 \phi_2 }{a_1 (b^2-\phi^2_1+\phi_2^2)-2 a_2 \phi_1 \phi_2 } \, ,
\end{equation}
can be integrated. Indeed, a family of solutions describing circumferences is obtained in the plane, which in terms of $(\lambda,\nu)$ can be written as
 \begin{equation}\label{eq:CircumferenceGeneral}
	\left(\phi_1-\Gamma \sin\nu\right)^2+\left(\phi_2+\Gamma \cos\nu\right)^2=\Gamma^2+ b^2\,,
\end{equation}
 where every member is labeled by the constant of integration $\Gamma\in\mathbb{R}$. Notice that the circumference centered at the origin $\Gamma=0$ is always a solution, but as the value of the constant $|\Gamma|$ increases, the center of the circumference will move depending on the angle $\nu$. In fact, it is straightforward to prove that solutions in any rotated coordinate system can be obtained by rotating the solutions obtained for the particular case with $a_2=0$ in \eqref{eq:SuperpotentialModel2} with no loss of generality. In order to obtain explicit solutions, let us restrict to this case with $a_1\equiv a$ and $a_2=0$
\[
W(\phi_1,\phi_2) = \frac{a \, \phi^1}{b^2+\phi_1^2+\phi_2^2} \,,
\]
for which the potential can also be written more compactly as 
\begin{equation}\label{eq:PotentialMod2PlanoRotado}
    V(\phi_1,\phi_2) =\frac{1}{2} a^2 \frac{ (\phi_1^2 + \phi_2^2 +b^2)^2 - 4 b^2 \phi_1^2}{(\phi_1^2 +\phi_2^2+b^2)^4}\,.
\end{equation}
Note that the discrete vacua are now located at points that lie in the $\phi_1-$axis
\[
{\cal M} = \{ v^1=(-b,0) \,\, , \,\, v^2=(b,0) ,  \,\, v_{\infty}\} \,.
\]
From the first order differential equations \eqref{eq:FirstOrderEqsModelo2a} and \eqref{eq:FirstOrderEqsModelo2b} for this case the following solutions are identified:

\begin{itemize}
    \item \textbf{Singular $\Phi_1-$kinks}: Imposing the trial orbit $\phi_2=0$ in the first order equations \eqref{eq:FirstOrderEqsModelo2a} and \eqref{eq:FirstOrderEqsModelo2b} in the plane the following solutions in implicit form are obtained for the $\phi^1$ component
    \begin{align}
        &-\phi_1+2 b \arctanh\frac{\phi_1}{b}=(-1)^{\epsilon} \, a(x-x_0) \qquad \text{if} \qquad |\phi^1|<b\\
        &-\phi_1+2 b \arccoth\frac{\phi_1}{b}=(-1)^{\epsilon} \, a(x-x_0) \qquad \text{if} \qquad |\phi^1|>b\,.
    \end{align}
    These equations represent three types of solutions. While one of them asymptotically connects vacua $v^1$ and $v^2$, the other two link each vacua $v^1$ and $v^2$ with the infinity. 

    \item \textbf{Families of Kinks}: The integration of the first order differential equation \eqref{eq:FirstOrderEqsModelo2a},  employing the orbit equation \eqref{eq:CircumferenceGeneral}, allows us to identify an implicit equation for members of this family of kinks
    \begin{align*}
&   \frac{C^2 \phi _2 \left(2 b \phi _2 -C F(\phi_2;C)\right)}{2 b F(\phi_2;C)+C \phi _2}+C \left(8 b^2+C^2\right) \arctan\left(\frac{F(\phi_2;C)}{\phi _2}\right)+\\
& +4 b^3 \log \left|\frac{2 b F(\phi_2;C)}{\phi _2}+C\right|= \pm \, 2 a \left(x-x_0\right) \,,
\end{align*}

where $C\in \mathbb{R}$ and $F(\phi_2;C)=\sqrt{b^2-\phi _2 \left(C+\phi _2\right)}-b$. Consequently, $\phi_1(x)$ is also determined via the orbit equation \eqref{eq:CircumferenceGeneral}.
\end{itemize}

Similarly to the previous section, a Sigma model with action \eqref{eq:actionMod1Esfera} shall be constructed in stereographic coordinates via \eqref{eq:PotentialRelation} so that the transferred solutions are solutions. Without any loss of generality, let us deform the rotated model \eqref{eq:PotentialMod2PlanoRotado}, leading to the potential function on the sphere
\begin{equation}\label{eq:Mod2EsferaRotado}
    \widetilde{V}(\psi_1,\psi_2) = \frac{1}{8} a^2 (\psi_1^2 + \psi_2^2 +1)^2 \frac{ (\psi_1^2 + \psi_2^2 +b^2)^2 - 4 b^2 \psi_1^2}{(\psi_1^2 +\psi_2^2+b^2)^4}\,.
\end{equation}
It is worth highlighting that even though in the model \eqref{eq:Mod2EsferaRotado} no dilation is considered, transformations in the parameters $a\rightarrow\mu^3 a$ and $b\rightarrow\mu \, b$ recover the dilated case. Therefore, studying the non-dilated case $\mu=1$ allows us to find solutions for the dilated one. Notice that, unlike in the model on the sphere \eqref{eq:Mod1PotEsf} described in Section \ref{sec:ExampleA}, the potential \eqref{eq:Mod2EsferaRotado} is well-defined at $\psi_1^2+\psi_2^2\rightarrow\infty$, which corresponds to the North Pole of the sphere. Moreover, the limit is finite
\[
\lim_{\psi_1^2 + \psi_2^2 \rightarrow \infty} \widetilde{V}(\psi_1,\psi_2)  = \frac{1}{8} a^2\,.
\]
This opens the possibility for the projection singular point to become a ``passing point'' of kink orbits.  Even though the original potential $V(\phi)$ vanishes at infinities, the value of the potential $\widetilde{V}(\psi)$ at the north pole does not vanish. Therefore, solutions in the plane that tend to the vacua at infinity will be connected by this point. This is evident when the vacuum manifold of this potential \eqref{eq:Mod2EsferaRotado} is identified:  
\[
{\cal M}|_{S^2} = \Big\{ \Big(-\frac{2b}{1+b^2}, 0, \frac{b^2-1}{b^2+1} \Big) , \Big(\frac{2b}{1+b^2}, 0, \frac{b^2-1}{b^2+1} \Big) \Big\}\,,
\]
where the original $v_{\infty}$ of the plane, now corresponding to the north pole, have not been inherited. Notice that the two vacua that arise are antipodal points on the sphere only if $b=\pm1$, when they lie in the equator.

Static solutions for this new model on the sphere can be found by repeating the previously described procedure. While the explicit reparametrization may not be analytically available in general, an implicit solution for the transferred kinks can be identified by solving Bogomol'nyi equations for the new superpotential on the sphere:
\begin{equation}
    \widetilde{W}(\psi_1,\psi_2) = \frac{a \, \psi^1}{b^2+\psi_1^2+\psi_2^2} \,.
\end{equation}
In this particular case, the first order equations give rise to two types of solutions, which are described below:

\begin{itemize}
    \item \textbf{Singular $\Psi-$kinks}: 
 Imposing the trial orbit $\psi_2=0$ in Bogomol'nyi equations leads to two different implicit relations depending on the value of $|\psi_1|$ along solutions. On one hand, when $|\psi_1|<1$ one obtains
\[
\frac{(1-b^2)^2\psi_1}{2(b^2+1)(1+\psi_1^2)} + \frac{(b^2-1)(b^4+6 b^2+1)}{2(b^2+1)^2} \, \arctan \psi_1 + \frac{4 b^3 \arctanh \frac{\psi_1}{b}}{(b^2+1)^2} = (-1)^{\epsilon} \, \frac{a}{4} \,(\tilde{x}-\tilde{x}_0)\,,
\]
while when $|\psi_1|>1$ another implicit relation is identified 
\[
\frac{(1-b^2)^2\psi_1}{2(b^2+1)(1+\psi_1^2)} + \frac{(b^2-1)(b^4+6 b^2+1)}{2(b^2+1)^2} \, \arctan \psi_1 + \frac{4 b^3 \arccoth \frac{\psi_1}{b}}{(b^2+1)^2} = (-1)^{\epsilon} \, \frac{a}{4} \,(\tilde{x}-\tilde{x}_0)\,.
\]
These solutions can be significantly reduced for the case where the vacua on the sphere are antipodal $b=\pm1$ to the explicit expression
    \begin{align}
        &\psi_1(\tilde{x}) =(-1)^{\epsilon}\, \tanh \frac{a (\tilde{x}-\tilde{x}_0)}{4}  \qquad \text{if} \qquad |\psi^1|<1\label{eq:Mod2Solucionb1a}\\
        &\psi_1(\tilde{x}) =(-1)^{\epsilon}\, \coth \frac{a (\tilde{x}-\tilde{x}_0)}{4}  \qquad \text{if} \qquad |\psi^1|>1\,.\label{eq:Mod2Solucionb1b}
    \end{align}
Notice that in this particular case, equation \eqref{eq:GeneralReparemetrisation} allows us to identify the relation between the new spatial parameter $\tilde{x}$ and the original $x$:

 \begin{equation*}
       x=x_0+ (\tilde{x}-\tilde{x}_0)+ \frac{4 \,  \omega}{a} \, {\rm Gd}\left(\frac{a (\tilde{x}-\tilde{x}_0)}{2}\right)+\frac{2}{a} \tanh\left(\frac{a (\tilde{x}-\tilde{x}_0)}{2}\right) \,,
    \end{equation*}
where $\omega$ is a constant with value $\omega=1$ when $|\psi_1|<1$ and $\omega=-1$ when $|\psi_1|>1$, see Figure \ref{figure:Model2SimpleRepara}.
\begin{figure}[h]
	\centerline{\includegraphics[height=5cm]{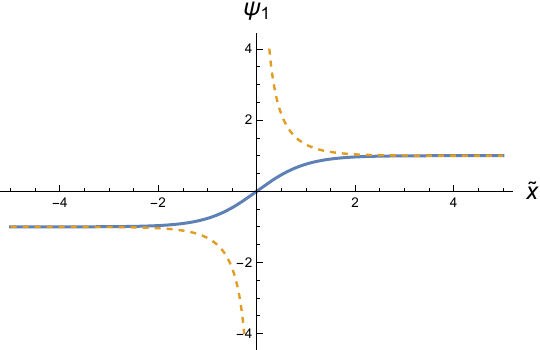} \hspace{0.3cm} \includegraphics[height=4.75cm]{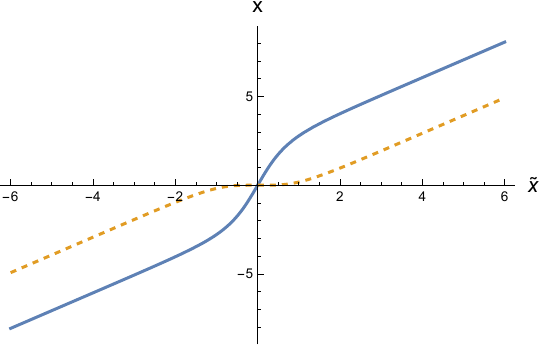}}
	\caption{\small \small Solutions \eqref{eq:Mod2Solucionb1a} and \eqref{eq:Mod2Solucionb1b} are shown on the left in blue thick line and in dashed golden line respectively and the reparametrization $x(\tilde{x})$ is depicted on the right. Particular values of parameters $\epsilon=0$, $a=4$ and $\tilde{x}_0=0$ are chosen for the representation of solutions. It is worth noticing how each one corresponds to a solution that crosses a different pole in the sphere. It is also worth noting that the derivative of the reparametrization with $\omega=-1$ vanishes at zero.}\label{figure:Model2SimpleRepara}
\end{figure}
In these solutions, vacua $v^1$ and $v^2$ are connected by two different pieces of the maximal circle that crosses both points in the sphere, see Figure \ref{figure:Model2SimpleRepara}. Note that one of these solutions crosses the south pole, located at $(0,0)$ in the stereographic plane, while the other crosses the north pole as a passing point. It should be stressed that  another chart that includes the north pole and excludes the south pole could have been considered. Indeed, it is straightforward to prove that the potential and the equations that arise in this case are equivalent to potential \eqref{eq:Mod2EsferaRotado} and its corresponding Bogomol'nyi equations. Therefore, studying the solutions in this chart is enough. 

On the other hand, the energy density for both $\Psi-$kinks when $b=1$ is identical and equal to 
\begin{equation}
    \tilde{\varepsilon}(\tilde{x})=\frac{a^2}{4} \sech^2\frac{a (\tilde{x}-\tilde{x}_0)}{2}\,,
\end{equation}
which corresponds to a single peak as in the singular kinks in the first model, see Figure \ref{figure:SGR2Densidades}. Lastly, the energy of all these singular kinks can be easily obtained from equation \eqref{eq:EnergyBPS}, which reads for both solutions
\[
\widetilde{E}[\Psi] = \frac{a}{b}\,.
\]
Note that for dilated kinks, transformations $a\rightarrow\mu^3 a$ and $b\rightarrow\mu \, b$ result in a global factor $\mu^2$ appearing in the energy as expected.

\item \textbf{Families of kinks $\Sigma$}: By construction, the form of the orbit flow equation \eqref{eq:OrbitEquationModel2} is conserved in the deformation. This leads to kink orbits describing circles in the stereographic plane with center $(0,-\frac{1}{2} \gamma)$ and radius $R^2= b^2 + \frac{1}{4} \gamma^2$ 
\begin{equation}\label{eq:OrbitaAlonso}
	 \psi_1^2+\Big(\psi_2+\frac{1}{2} \gamma \Big)^2 = b^2 + \frac{1}{4} \gamma^2 \hspace{0.4cm},\hspace{0.4cm} \gamma\in \mathbb{R}\,.
\end{equation}
 In summary, every circle orbit involves two kinks (and two antikinks) asymptotically joining the two vacua on the sphere, see Figure \ref{figure:Modelo2Orbitas}. Clearly, since the same two vacua are asymptotically connected by all these kinks, the energy \eqref{eq:EnergyBPS} of every member of this family of kinks is identical
\[
\widetilde{E}[\Sigma] = \frac{a}{b}=\widetilde{E}[\Psi]\,,
\]
and identical to that of the singular kinks. Both in the Euclidean plane and the sphere solutions in the limit $|\gamma|\rightarrow \infty$ tend to both singular kinks. However, in the case of the sphere, when the singular kinks that tend to infinity in the plane are transferred, these are glued at the north pole, where the potential tends to a positive value. This is, the singular kink that crosses the north pole is formed. Moreover, substituting the orbit equation \eqref{eq:OrbitaAlonso} into the Bogomol'nyi equations allows us to derive an implicit expression for $\psi_2(x)$ 
\begin{align*}
    &\frac{\left(b^2-1\right) \gamma  \Lambda(\gamma)}{\Delta
   ^3} \arctan\left(\frac{\left(b^2+1\right) f\left(\psi
   _2,\gamma \right)-b \left(b^2+\gamma  \psi _2+1\right)}{\Delta  \psi _2}\right)-4 b^3 \log \left| \gamma +\frac{2 b \left(b-f\left(\psi _2,\gamma \right)\right)}{\psi
   _2}\right| \\
    &-\frac{\left(b^2-1\right)^2 \gamma ^2 \psi _2 \left(\left(\gamma -b^2 \gamma \right)
   f\left(\psi _2,\gamma \right)+b \psi _2 \left(2 b^2+\gamma ^2+2\right)+b
   \left(b^2-1\right) \gamma \right)}{\Delta ^2 \left(b^2+\gamma  \psi _2+1\right) \left(2 b^2+\gamma  \psi_2 -2
   b f\left(\psi _2,\gamma \right)\right)}\\
   &= \pm \, \frac{a}{2} (1+b^2)^2 (\tilde{x}-\tilde{x}_0) \,,
\end{align*}
where $f(\psi_2;\gamma)=\sqrt{b^2+\psi _2 \left(\gamma -\psi _2\right)}$, $\Delta(\gamma)=\sqrt{b^4+2 b^2+\gamma ^2+1}$ and $\Lambda(\gamma)=8 b^6+b^4 \left(\gamma ^2+16\right)+b^2 \left(6
   \gamma ^2+8\right)+\gamma ^2$. For the antipodal case $b=1$ these equations can be further simplified for every member $\gamma$ and explicit expressions for the kinks can be identified
\begin{eqnarray*}
\psi_1(\tilde{x};\gamma) &=& (-1)^{\epsilon_3} \left| \frac{(3 + \gamma^2) \cosh\hat{x} + (5 + \gamma^2) \sinh\hat{x}}{2 \gamma (-1)^{\epsilon_2} + (5 + \gamma^2) \cosh\hat{x}+ (3 + \gamma^2) \sinh\hat{x}} \right|\,,
 \\
\psi_2(\tilde{x};\gamma) &=& -\frac{4 (-1)^{\epsilon_2}}{2 \gamma (-1)^{\epsilon_2} + (5 + \gamma^2) \cosh\hat{x} + (3 + \gamma^2) \sinh\hat{x}}\,,
\end{eqnarray*}

where the sign in $\hat{x}=\pm \frac{1}{2} a (\tilde{x} - \tilde{x}_0)$ distinguishes between kink and antikink and $\epsilon_1,\epsilon_2,\epsilon_3=0,1$.

\begin{figure}[ht]
	\centerline{\includegraphics[height=5cm]{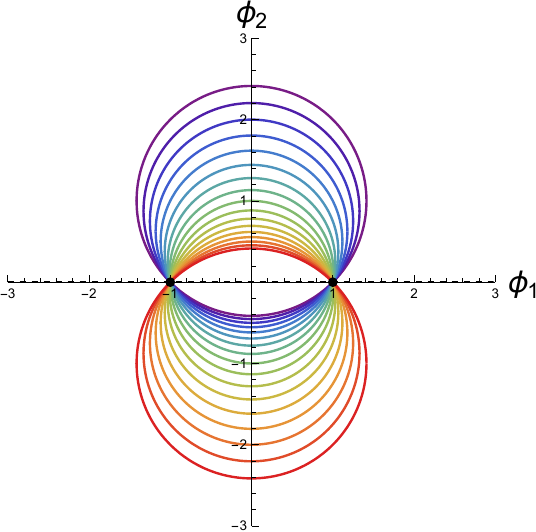} \hspace{0.6cm} \includegraphics[height=5cm]{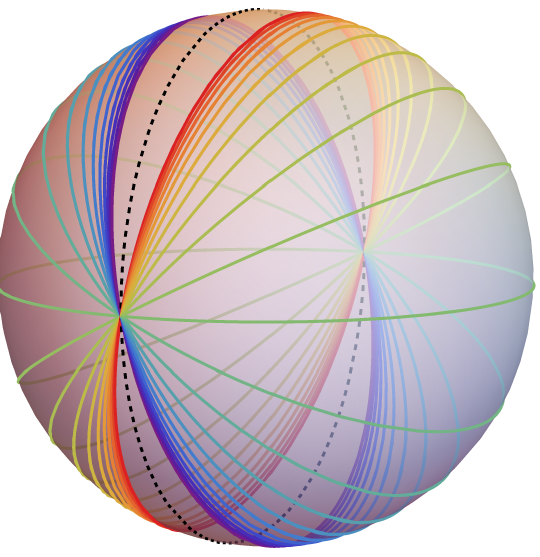}}
	\caption{\small Members of the family of kinks of the model \eqref{eq:PotentialMod2PlanoRotado} in the plane (left) and the deformed family of kinks for the model on the sphere \eqref{eq:Mod2EsferaRotado} (right) are respectively depicted with $b=1$. These solutions correspond to pieces of circumferences whose radii increase as $|\gamma|$ increases. Notice that the limit case with infinite radius corresponds to singular $\Psi-$kinks. In the case of the model on the sphere, one of these singular kinks crosses the south pole while the other crosses the north pole.}\label{figure:Modelo2Orbitas}
\end{figure}
Note that limits $\gamma\rightarrow\infty$ and $\gamma\rightarrow-\infty$ produce different solutions. While $\gamma\rightarrow\infty$ corresponds to the singular kink that crosses the south pole $(\tanh{\frac{a (\tilde{x}-\tilde{x}_0)}{4}},0)$, the limit $\gamma\rightarrow-\infty$ corresponds to the one that crosses the north pole $(\coth{\frac{a (\tilde{x}-\tilde{x}_0)}{4}},0)$.

\end{itemize}

It is worth highlighting that, even though the orbit equations of the original and deformed models are identical, the identification of points at infinities during the deformation does not suffice to find a transferred kink crossing the north pole. Indeed, even if limits of solutions tend to infinities in the original model \eqref{eq:PotentialMod2PlanoRotado}, the fact that allows the north pole to function as a passing point is that the value of the deformed potential is positive at this point.

\subsection{Example C}
 In this section the Mercator projection of the sphere onto the cylinder, see Appendix, is explored to deform a Sigma model in the plane to another on the sphere.  Since the sphere is also locally conformally flat in these coordinates, the same procedure from previous sections of construction of the potential will be followed. In particular, let us consider the model in the plane with the following superpotential written in Cartesian coordinates $\{\phi^1,\phi^2\}$
\begin{equation}
	W(\phi_1,\phi_2)=\sin\frac{\phi_2}{2} ~ e^{-\phi_1^2}\,,
\end{equation}
for which a potential that vanishes at $\phi_1\rightarrow\pm \infty$, $\forall \phi_2$ is constructed
\begin{equation}\label{eq:ModPotentialPlane2}
    V(\phi_1,\phi_2)=\frac{1}{8} e^{-2\phi_1^2} \left[ \cos^2\frac{\phi_2}{2}+16 \, \phi_1^2 \, \sin^2\frac{\phi_2}{2} \right]\,.
\end{equation}
Given the periodicity of this potential in the field $\phi_2$, it  presents an infinite number of vacua at points
\begin{equation*}
    \mathcal{M}=\left\{(0,\pi + 2 \pi n)\right\} \,,
\end{equation*}
where $n\in \mathbb{Z}$. Moreover, potential \eqref{eq:ModPotentialPlane2} tends to zero as $\phi_1\rightarrow \pm\infty$, which corresponds to points that are mapped to the north and south poles via the Mercator antiprojection. Several vacua can be seen depicted in Figure \ref{figure:Mod3Sing2} (left). On the other hand, this superpotential gives rise to the following Bogomol'ny equations
\begin{align} 
    \frac{d\phi_1}{d x}=&(-1)^{\epsilon+1} \, 2 \, \phi_1 e^{-\phi_1^2}  \, \sin\frac{\phi_2}{2}\,,\label{eq:BogomolnyiModel32}\\
    \frac{d\phi_2}{d x}=&(-1)^{\epsilon} \frac{1}{2} \,  e^{-\phi_1^2}  \, \cos\frac{\phi_2}{2} \,,\label{eq:BogomolnyiModel3bA2} 
\end{align}
where once again $\epsilon=0,1$. The trial orbit method allows us to identify the different arising singular kinks in the plane, see Figure \ref{figure:Mod3Sing2}. The orbit equation for the families of kinks can be easily derived by the integration of the orbit flow equation:
\begin{equation*}
    \frac{d\phi_1}{d\phi_2}=  -4 \,\phi_1\, \tan\frac{\phi_2}{2}\,.
\end{equation*}
leading to a family of orbits labeled by the constant $K>0$
\begin{equation}
    \phi_1= \pm K \cos^{8}\frac{\phi_2}{2}\,.
\end{equation}
These families of kinks, that emerge in the plane in regions delimited by the parallel singular kinks, are represented for the central regions in Figure \ref{figure:Mod3Families2}.

On the other hand, a deformation of this model in the plane can be performed, transferring kinks of this model to another on the sphere making use of the Mercator projection. For convenience, let us not perform any dilation in the transference. Moreover, the introduced coordinates on the sphere $\{\psi_1,\psi_2\}$ will be denoted as $\{\Omega,\alpha\}$ for the Mercator projection \eqref{eq:CoordinatesMercator}, where $\Omega\in(-\infty,\infty)$ and $\alpha\in[-\pi,\pi)$. The previously described procedure then leads in this last case to a model characterized by the following superpotential on the sphere
\begin{equation}
	W(\Omega,\alpha)=\sin\frac{\alpha}{2} ~ e^{-\Omega^2}\,,
\end{equation}
for which the corresponding potential is periodic in the angle $\alpha$ on the sphere
\begin{equation}
    \widetilde{V}(\Omega,\alpha)=\frac{\cosh^2{\Omega}}{8} e^{-2 \Omega^2} \left[ \cos^2\frac{\alpha}{2}+16 ~ \Omega^2 \, \sin^2{\frac{\alpha}{2}}\right]\,.
\end{equation}
It is worth highlighting that, as it was shown in \cite{AlonsoIzquierdo2022}, the non-periodicity of the superpotential in the angular variable allows us to identify closed orbit kinks. In this section the potential tends to zero at the singular projection points, which implies that these points become vacua of the field theory on the sphere. Indeed, the limit of this potential at both south and north poles vanish 
\begin{equation*}
    \displaystyle\lim_{\Omega\rightarrow\pm\infty}\widetilde{V}(\Omega,\alpha)=0 \qquad \forall\alpha\,.
    \end{equation*}
In total, this potential function on the sphere presents three vacua which are aligned in a meridian
\begin{equation*}
    \mathcal{M}=\left\{(0,\pi),N,S\right\} \,,
\end{equation*}
where the north pole $N$ and south pole $S$ corresponds to limits $\Omega\rightarrow\infty$ and $\Omega\rightarrow-\infty$ respectively, see Figure \ref{figure:Mod3Sing2}. Note that the vacua that arise in the model in the plane are now identified in the transference to the sphere $(\Omega,\pi) \rightarrow(\Omega,-\pi)$. Indeed, while the model in the plane presents a vacuum point replicated every $2\pi$ in $\phi_2$, all these points are identified as the same point in the cylinder via the Mercator projection. Furthermore, points corresponding to $\phi_1\rightarrow\infty$ and $\phi_1\rightarrow-\infty$ are mapped via this projection to the north and south poles respectively. Notice that even if this identification is performed in the central area $\phi_2\in(-\pi,\pi)$, any other area of the plane could have been chosen to form the cylinder since solutions in the plane are $2\pi-$periodic in this coordinate. This exemplifies how the transference between manifolds can be chosen to generate a decrease in the number of vacua in the deformed model, see Figure \ref{figure:Mod3Sing2}. 

The form of the deformed superpotential in these coordinates leads to the following Bogomol'nyi equations
\begin{align} 
    \frac{d\Omega}{d\tilde{x}}=& (-1)^{\epsilon+1} \, 2\, \Omega \, \cosh^2{\Omega} \, e^{-\Omega^2} ~ \sin\frac{\alpha}{2}\,,\label{eq:BogomolnyiModel32}\\
    \frac{d\alpha}{d \tilde{x}}=& (-1)^{\epsilon} \frac{\cosh^2{\Omega}}{2} \, e^{-\Omega^2}  ~ \cos\frac{\alpha}{2} \,.\label{eq:BogomolnyiModel3b2} 
\end{align}

For this model on the sphere, three types of singular kinks, for which one of the variables is constant along the orbit, are identified. Moreover, apart from these singular kinks, a whole family of kink orbits is analytically obtained. These solutions are described below.

\begin{itemize}
    \item \textbf{Singular} $\alpha-$\textbf{kinks}: When the trial orbit with constant variable $\Omega=0$ is imposed in Bogomol'nyi equations \eqref{eq:BogomolnyiModel32} and \eqref{eq:BogomolnyiModel3b2}, the following expression for the singular $\alpha-$kink is obtained
    \begin{equation}
       \alpha(\tilde{x})=(-1)^{\epsilon} \,2  \, {\rm Gd}\left(\frac{\tilde{x}-\tilde{x}_0}{4}\right) \,,
    \end{equation}
    where $x_0\in\mathbb{R}$ is the center of the kink. It is worth noting that these solutions are parallels in the sphere, describing non-topological kinks, see Figure \ref{figure:Mod3Sing2}.
    \item \textbf{Singular} $\Omega-$\textbf{kinks}: When the trial orbits $\alpha=\pm\pi$ with $\epsilon_2=0,1$ are imposed in the Bogomol'nyi equations \eqref{eq:BogomolnyiModel32} and \eqref{eq:BogomolnyiModel3b2}, quadratures are not solvable. However, from these equations it follows that for both values $\alpha=\pm \pi$ topological singular kinks appear asymptotically connecting adjacent vacua along a meridian, see Figure \ref{figure:Mod3Sing2}. For $\alpha=\pi$ these two singular kinks will be labeled as $\Omega_{-}$ and $\Omega_{+}$ for the orbit with $\Omega<0$ and $\Omega>0$ respectively. Complementarily, for $\alpha=-\pi$ these kinks will be denoted as $\overline{\Omega}_{-}$ and $\overline{\Omega}_{+}$, even though these are identified as $\Omega_{-}$ and $\Omega_{+}$ in the sphere. It is worth noticing that while this identification is performed in the sphere, these are different kinks in the original model \eqref{eq:ModPotentialPlane2} in the plane.

\begin{figure}[h]
	\centerline{\includegraphics[height=4cm]{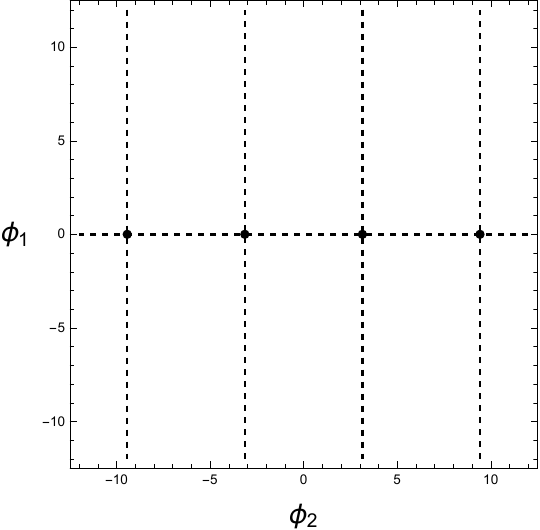} \hspace{0.4cm}
		\includegraphics[height=4cm]{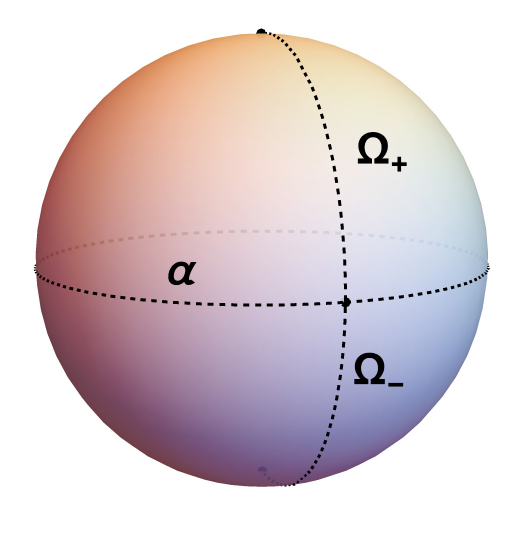}}
	\caption{\small Singular kinks in the plane and sphere respectively. The singular kinks that are replicated in the plane every $\phi_2\rightarrow \phi_2+2\pi$ are identified in the sphere, resulting in one singular $\alpha-$kink and two $\Omega-$kinks in the sphere. Notice that while $\alpha-$kinks are closed orbit kinks revolving the sphere, $\Omega-$kinks are fragments of a meridian.}\label{figure:Mod3Sing2}
\end{figure}

\item \textbf{Family of kinks}: Once more Bogomol'nyi equations \eqref{eq:BogomolnyiModel32} and \eqref{eq:BogomolnyiModel3b2} cannot be directly integrated. Since this deformation procedure preserves the form of the orbit equation, it provides a family of orbits labeled by the constant $K>0$
\begin{equation}
    \Omega=\pm K \cos^{8}\frac{\alpha}{2}\,,
\end{equation}
see Figure \ref{figure:Mod3Families2}. While this expression is formally identical to that for kinks in Cartesian coordinates on the plane, here the presence of the angle $\alpha$ restricts the solution on the sphere to one of the bands in the plane. Furthermore, it is worth noticing that this family of kinks contains only closed orbit kinks. On the other hand, this family of kinks can be decomposed into two subfamilies of kinks, which shall be denoted as $\Sigma_{+}$ and $\Sigma_{-}$ for those with $\Omega>0$ and $\Omega<0$ respectively.

\begin{figure}[h]
	\centerline{\includegraphics[height=4cm]{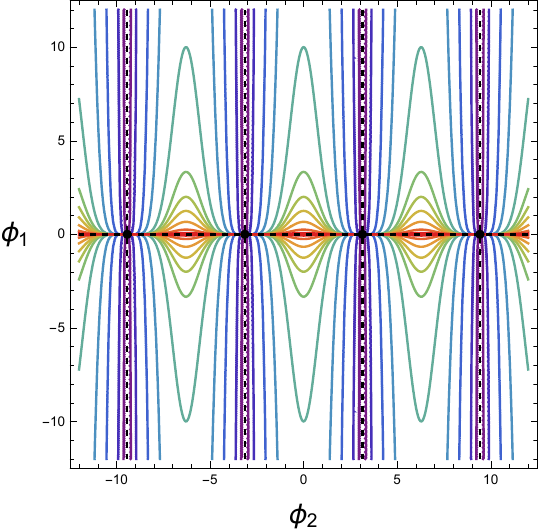} \hspace{0.4cm}
		\includegraphics[height=4cm]{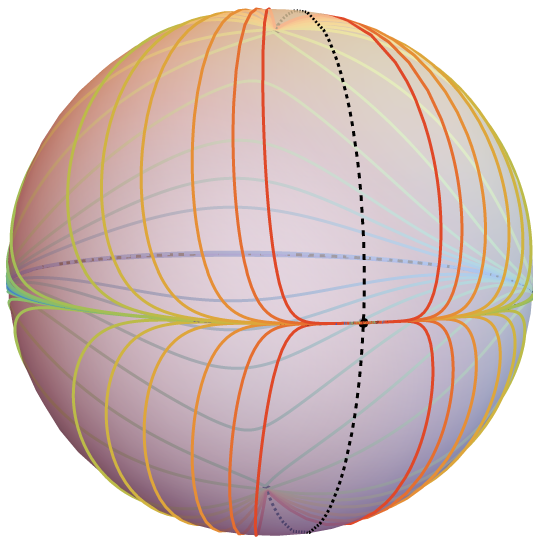}}
	\caption{\small Orbits of families of kinks in the plane and in the sphere respectively. Two subfamilies of kinks emerge replicated in the plane $\phi_2\rightarrow\phi_2+2\pi$ as a result of the integration of the orbit flow equation. Kinks in every band in the plane are identified in the transference and densely fill the two different regions of the sphere delimited by the singular $\alpha-$kink.}\label{figure:Mod3Families2}
\end{figure}

The energy of all these kinks, singular and members of the families, can be easily computed and an energy sum rule is identified: 
\begin{equation*}
    \widetilde{E}[\Sigma_{\pm}]=\widetilde{E}[\alpha]=2\,, \qquad \widetilde{E}[\Omega_{\pm}]=1\,, \qquad \widetilde{E}[\Sigma_{\pm}]=\widetilde{E}[\Omega_{\pm}]+\widetilde{E}[\overline{\Omega}_{\pm}]=\widetilde{E}[\alpha] \,.
\end{equation*}
This is consistent with the limits of this family for the parameter $K$. When $K\rightarrow\infty$ the $\alpha-$kink with $\Omega=0$ is recovered. On the other hand, for $K\rightarrow 0$ the $\Omega-$kinks with $\alpha=\pm \pi$ are obtained as limits.

   \end{itemize}

   This procedure of deformation not only allows us to transfer kinks to new models by reparametrization of solutions, but also allows us to transform topological kinks into non-topological kinks given the change of target manifold. Notice that once more, vacua at infinities that were reached only as a limit in the plane are now part of the target manifold.

\section{Conclusions}

In this work deformations extensively studied in the literature \cite{Afonso2007, Almeida2004, AlonsoIzquierdo2013, Bazeia2002, Bazeia2006a, Bazeia2006b, Bazeia2011, Bazeia2013, Bazeia2017, Chumbes2010, Cruz2009} have been generalized. In particular, deformations of Sigma models in the plane to others on the sphere $\mathbb{S}^2$ have been performed. The transference of solutions between manifolds requires the choice of a chart on each manifold with which kink orbits can be related. Moreover, a relation between the parametrizations of these curves on each chart is also considered. The presented formalism allows us, on one hand, to perform deformations between the plane and any conformally flat manifold by making the reparametrization absorb the change in the metric. Indeed, the orbit equations for both charts will be formally identical. This also implies that different solutions will give rise to different parametrizations $\tilde{x}(x)$ for which the solutions are known. The explicit identification of a solution may not be possible if the reparametrization does not have an analytic inverse. Nevertheless, the deformed orbits are still known.

On the other hand, the difference in geometry between the plane and the sphere makes the profiles of the transferred solutions be very different. Moreover, the fact that the sphere is compact may alter the topological sectors of the deformed kink variety. In the first example of deformation, example A, a sine-Gordon model in the plane is deformed, transferring solutions to the sphere by means of the inverse of the stereographic projection. While the potential at infinities of the plane is not defined, the deformed potential on the sphere presents a singularity at the north pole. This prevents deformed solutions from reaching the north pole in the sphere. 

In the second scenario, the example B, a rational superpotential is considered. This leads to a model with two discrete vacua in the plane, but also a continuous set of vacua at infinity $v_{\infty}$. When the deformed Sigma model on the sphere is constructed in this case, the set of vacua $v_{\infty}$ is not inherited. Instead, the deformed potential on the sphere has a non-zero finite value at the north pole. This allows us to identify a new kink orbit crossing the north pole, which in the plane were two pieces clearly different. 

Lastly, in example C the stereographic projection is replaced by the Mercator projection, where the sphere is projected to a cylinder. This implies that while the orbits of the original and deformed models are identical, only a band in the plane is sent to the sphere given the angular identification of one of the coordinates $\alpha\rightarrow \alpha+2\pi$. As a consequence, some of the topological kinks in the plane are transformed into non-topological kinks that now revolve the sphere.

It is worth noticing that a mechanism for modulating the energy of the transferred kinks is introduced. By including a dilation in the relation between coordinates of each chart, the energy of the transferred kinks is directly affected by the parameter that controls such a dilation. Indeed, this parameter controls the height of the peaks of the potential and therefore the energy of the emerging kinks. Furthermore, since symmetrical dilations scale the energy of all kinks equally, the energy sum rules present in the model in plane are inherited in the model on the sphere.  Lastly, deformations between other Riemannian manifolds could be considered. For instance, deformations between connected and non-simply connected target manifolds could entail interesting changes in the topological sectors of kinks.

\section*{Acknowledgements}

This research was funded by the Spanish Ministerio de Ciencia e Innovación (MCIN) with funding from the European Union NextGenerationEU
(PRTRC17.I1) and the Consejería de Educación, Junta de Castilla y León, through QCAYLE project, as well as grants PID2020-113406GB-I00 MTM
and PID2023-148409NB-I00 MTM funded by MCIN/AEI/10.13039/501100011033.

\appendix

\section{Stereographic and Mercator projections}
 \label{sec:appendixA}
In this appendix the coordinate systems on the sphere that are employed along this work are briefly presented. The fact that the sphere is locally conformally flat in both coordinate systems allows us to apply in both cases the deformation procedure described in section $2$. 
\begin{itemize}
    \item \textbf{Stereographic projection}:

 If the sphere of radius $R=1$ is embedded in $\mathbb{R}^3$, coordinates in the ambient space $(X, Y, Z)$ must satisfy:
\[
X^2+Y^2+Z^2=1\,.
\]
In particular, the stereographic projection $P_S:\mathbb{S}^2-\{ N\}\rightarrow \mathbb{R}^2$ that projects from the North Pole $N\equiv (0,0,1)$ onto the plane $Z=0$ shall be used. This relates the coordinates of both charts $P(X,Y,Z)=(\psi_1,\psi_2)$ where
\begin{equation}
\psi_1=\frac{X}{1-Z}\  ,\quad \psi_2=\frac{Y}{1-Z} \,.\label{stereo}
\end{equation}
Equivalently, the inverse map $P^{-1}$ will provide us with the coordinates on the sphere for the ambient space when the projective coordinates on the plane are known:
\begin{equation}
    (X,Y,Z)=\left(\frac{2\psi_1}{1+\psi_1^2+\psi_2^2},\frac{2\psi_2}{1+\psi_1^2+\psi_2^2},\frac{-1+\psi_1^2+\psi_2^2}{1+\psi_1^2+\psi_2^2}\right)\,.
\end{equation}
In these coordinates, the form of the metric tensor on the sphere clearly shows that it is locally conformally flat:
\begin{equation}
\left.ds^2\right|_{\mathbb{S}^2}=\frac{4 }{\left(1+\psi_1^2+\psi_2^2\right)^{2}} \left( d\psi_1\otimes d\psi_1+d\psi_2\otimes d\psi_2\right)\,.\label{metric1}
\end{equation}

\item \textbf{Mercator projection}:

Alternatively, one may consider projecting the sphere of unit radius onto other spaces. Another example of conformal transformation is the Mercator projection, where points of the sphere are projected onto a cylinder. Let us start with spherical coordinates on the sphere
\begin{equation*}
    X= \sin{\theta} \cos{\varphi}\,,\qquad  Y=\sin{\theta} \sin{\varphi}\,,\qquad Z=\cos{\theta}\,,
\end{equation*}
where $\theta\in(0,\pi)$ and $\varphi\in[0,2\pi)$. In these coordinates the sphere is not conformally flat as it is obvious when the metric tensor is computed
\begin{equation*}
    \left.ds^2\right|_{\mathbb{S}^2}= d\theta\otimes d\theta+ \sin^2{\theta} \, d\varphi\otimes d\varphi\,.
\end{equation*}
When a change of coordinates $(\theta,\varphi)\rightarrow (\Omega(\theta),\alpha(\varphi))$ as the following is performed 
\begin{equation}\label{eq:CoordinatesMercator}
    \Omega(\theta)=\log\left[\tan\left(\frac{\theta}{2}\right)\right]\,, \qquad \alpha(\varphi)=\varphi\,,
\end{equation}
  the form of the metric tensor reveals that this space is conformally flat in these coordinates
\begin{equation*}
    \left.ds^2\right|_{\mathbb{S}^2}= \sech^2\Omega \left[d\Omega\otimes d\Omega+d\alpha\otimes d\alpha\right]\,.
\end{equation*}
 This change of coordinates defines a projection of the sphere without poles onto the cylinder $P_M:\mathbb{S}^2-\{ N,S\}\rightarrow [-\pi,\pi)\times\mathbb{R}$, where the poles of the sphere are obtained now only as limits at infinities of the variable $\Omega\in(-\infty,\infty)$. 
\end{itemize}

\end{document}